\documentclass[notitlepage,reprint,superscriptaddress]{revtex4-2}

\usepackage{amsmath}
\usepackage{graphicx}
\usepackage{float}
\usepackage{xcolor}
\usepackage{subcaption}
\usepackage{caption}
\usepackage{bm, upgreek}
\usepackage{hyperref}
\usepackage{autonum}
\newcommand{\ub}{\bm{\upbeta}}
\newcommand{\uo}{\bm{\upomega}}
\newcommand{\B}{\mathbf{B}}
\newcommand{\E}{\mathbf{E}}
\newcommand{\D}[2]{\frac{\mathrm{d}#1}{\mathrm{d}#2}}
\newcommand{\dd}{\mathrm{d}}
\newcommand{\Rom}[1]{\MakeUppercase{\romannumeral #1}}

\begin{document}

\title{New method of probing an oscillating EDM induced by axionlike dark matter using an RF Wien Filter in storage rings}
\author{On Kim}
\email[Corresponding author, ]{bigstaron9@ibs.re.kr}
\affiliation{Center for Axion and Precision Physics Research, Institute for Basic Science, Daejeon 34051, Republic of Korea}
\author{Yannis K. Semertzidis}
\affiliation{Center for Axion and Precision Physics Research, Institute for Basic Science, Daejeon 34051, Republic of Korea}
\affiliation{Department of Physics, Korea Advanced Institute for Science and Technology, Daejeon 34141, Republic of Korea}

\begin{abstract}
	A hypothetical pseudo-scalar particle axion, which is an immediate result of the Peccei-Quinn solution to the strong CP problem, may couple to gluons and lead to an oscillating electric dipole moment (EDM) of fundamental particles. This paper proposes a novel method of probing the axion-induced oscillating EDM in storage rings, using a radiofrequency (RF) Wien Filter. The Wien Filter at the frequency of the sidebands of the axion and $g-2$ frequency, $f_\text{axion} \pm f_{g-2}$, generates a spin resonance in the presence of an oscillating EDM, as confirmed both by an analytical estimation of the spin equations and independently by simulation. A brief systematic study also shows that this method is unlikely to be limited by Wien Filter misalignment issues.
\end{abstract}

\maketitle

\section{Introduction}
The Peccei-Quinn solution to the strong CP problem requires the existence of a pseudo-scalar Goldstone boson called an \textit{axion}\cite{Peccei1977}. The axion is a strongly motivated particle beyond the Standard Model and is also a plausible candidate for cold dark matter\cite{Preskill1983, Abbott1983, Dine1983}. While most axion search experiments seek to observe the axion-photon interaction using the resonant cavity method, it has been suggested that axion-gluon coupling in the strong interactions may result in an electric dipole moment (EDM) from a hadron oscillating at the axion frequency\cite{Graham2011, Graham2013}. A non-zero EDM of a hadron would require $CP$-violation in strong interactions, so there have been many efforts to measure the EDM of neutrons\cite{Altarev1980, Ramsey1982, Abel2020} and plans to measure the EDM of protons and/or deuterons using a storage ring method\cite{Farley2004, Orlov2006, Anastassopoulos2016, Semertzidis2016}. Extending the experimental approach to the axion-induced oscillating EDM, a new axion-like dark matter search experiment was proposed using the storage ring method in the presence of an oscillating EDM\cite{Chang2019, Stephenson2020, Pretz2020}. Another recent study also proposed that the storage ring EDM method can be exploited to probe dark matter and dark energy\cite{Graham2021}.

The present paper also proposes using the storage ring method to search for axions, but with a different scheme. We introduce an RF Wien Filter (WF) and resonate the spin by applying the RF at the sidebands of the axion frequency and the $g-2$ frequency. The application of the WF to measure the EDM was studied in Ref. \cite{Morse2013}, but its target was a conventional static EDM, while the present study seeks to observe an oscillating EDM induced by an axion field. Also, by applying the WF at a frequency other than just $g-2$ frequency, the experiment can be freed from the severe systematics arising from beam and spin dynamics and WF misalignment issues.

\section{Spin Dynamics in Storage Rings}
The spin of a particle in a storage ring precesses as
\begin{align}\label{eqn:Spin}
	\D{\mathbf{S}}{t} = \uo_s \times \mathbf{S},
\end{align}
where the angular spin frequency $\uo_s$ is given by the Thomas-BMT equation\cite{Thomas1926, Bargmann1959}:
\begin{widetext}
\begin{equation}\label{eqn:TBMT}
\begin{split}
\uo_s = -\frac{q}{m} \bigg[ \left(G+\frac{1}{\gamma} \right) \B - G \frac{\gamma}{\gamma+1} (\ub \cdot \B) \ub
 - \left( G + \frac{1}{\gamma+1} \right) \frac{\ub \times \E}{c}
 + \frac{\eta}{2} \left( \frac{\E}{c} - \frac{\gamma}{\gamma+1} \left( \ub \cdot \frac{\E}{c} \right) \ub + \ub \times \B \right) \bigg],
\end{split}
\end{equation}
\end{widetext}
where $\ub = \mathbf{v}/c$ is the particle velocity vector.
Here $G \equiv (g-2)/2$ is a magnetic anomaly and $\eta$ is a unitless EDM that plays the same role as the $g$-factor in the magnetic dipole moment. The magnetic dipole moment $\mu$ and the electric dipole moment $d$ can be written in forms:
\begin{align} \label{eq:definition_mu_and_d}
	\mu = g \frac{q}{2m} S, \qquad d = \eta \frac{q}{2mc} S,
\end{align}
where $q$, $m$ and $S$ are the charge, mass and spin of the particle, respectively.

We work in an accelerator coordinate system $(x, y, s)$, where $x$ is an in-plane radial distance from the design orbit, $y$ is an out-of-plane vertical distance from the center and $s$ is along the arc length of the storage ring. The spin angular frequency has only transverse components $(x, y)$ under a homogeneous and uniform vertical dipole magnetic field and/or radial electric field in the storage ring.
Furthermore, we employ a paraxial accelerator approximation $\ub \cdot \B = \ub \cdot \E = 0$ only for the analytical estimations, but those terms are kept in the numerical simulations and the results were consistent. Accordingly, the radial and vertical components of the spin angular frequency are given by
\begin{equation}\label{eqn:omega_xy}
\begin{split}
	\omega_{sx} &= -\frac{q}{2m} \eta(t) \left( \frac{E_x}{c} - \beta B_y \right), \\
	\omega_{sy} &= -\frac{q}{m} \left[ \left( G + \frac{1}{\gamma} \right) B_y - \left( G + \frac{1}{\gamma+1} \right) \frac{\beta E_x}{c} \right].
\end{split}
\end{equation}
Here we used the time-varying EDM for a hadron as a result of axion-gluon coupling:
\begin{align} \label{eqn:eta}
	\eta(t) = \eta_\text{DC} + \eta_\text{AC} \cos(\omega_\text{axion} t + \phi_\text{axion}).
\end{align}
For the analytical calculations of beam and spin dynamics for non-reference particles with non-zero EDM in storage rings, see Refs. \cite{Silenko2006, Fukuyama2013, Abusaif2021}.

\section{Spin Resonance with RF Wien Filter}
The RF Wien Filter (WF) is a perfect candidate to drive the spin resonance without affecting the beam betatron oscillations, since it exerts no Lorentz force on particles with a specific momentum. The EDM term $\eta(t)$ only contributes to the radial component of the spin angular frequency, so we set the electromagnetic field of the WF as follows.
\begin{equation}\label{eqn:WFfields}
\begin{split}
	\E_\text{WF} &= E_0^\text{WF} \cos(\omega_\text{WF} t + \phi_\text{WF}) \hat{e}_x, \\
	\B_\text{WF} &= \frac{E_0^\text{WF}}{\beta c} \cos(\omega_\text{WF} t + \phi_\text{WF}) \hat{e}_y
\end{split}
\end{equation}
An artificial spin resonance driven by a WF in the presence of a static non-zero EDM has been well studied\cite{Morse2013, Saleev2017, Slim2016, Rathmann2020}. We extend this idea to the oscillating component of the EDM. It is intuitive to expect that the vertical spin component will accumulate in one direction when the oscillation frequency of the EDM, $\omega_\text{axion}$, is the same as the WF frequency. Actually, it turns out that it is resonant with the sidebands of the axion and $g-2$ frequency: $\omega_\text{WF} = \omega_{g-2} \pm \omega_\text{axion}$.
It is also true that its sidebands with a cyclotron frequency, for instance $\omega_c - (\omega_{g-2} \pm \omega_\text{axion})$, are also resonance frequencies, because the WF is normally located in a specific position in the azimuth and the coherent spin motion with respect to the WF will include aliased Fourier components. However, in this paper we will assume the WF is continuously located in the azimuth to simplify the spin equation and solve it analytically.

To see the resonance condition $\omega_\text{WF} = \omega_{g-2} \pm \omega_\text{axion}$ explicitly, let us solve the spin equations for a reference particle that travels the storage ring in the reference orbit. Let $E_0$ and $B_0$ be the magnitudes of a constant radial electric field and a vertical magnetic field, respectively, needed to store the particle in the storage ring. Substituting the spin components in Eq. \eqref{eqn:Spin} with Eq. \eqref{eqn:omega_xy} and using the electromagnetic field $\E = E_0 \hat{e}_x + \E_\text{WF}$ and $\B = B_0 \hat{e}_y + \B_\text{WF}$ from Eq. \eqref{eqn:WFfields} yields
\begin{align}
	\dot{S}_x &= -(\omega_{g-2} + \Omega_\text{WF}(t) ) S_s, \\
	\dot{S}_y &= -\omega_\eta(t) S_s, \label{eqn:Sy} \\
	\dot{S}_s &= \omega_\eta(t) S_y + (\omega_{g-2} + \Omega_\text{WF}(t) ) S_x,
\end{align}
where
\begin{align}
	\omega_{g-2} &= \frac{q}{m} \left[ G B_0 - \left( G - \frac{1}{\gamma^2-1} \right) \frac{E_0}{c} \right]
\end{align}
is the $g-2$ frequency, and
\begin{align}
	\Omega_\text{WF} (t) &= \frac{q}{m} \frac{G+1}{\gamma^2} \frac{E_0^\text{WF}}{\beta c} \cos(\omega_\text{WF} t + \phi_\text{WF}) \\
	&\equiv a_\text{WF} \cos(\omega_\text{WF} t + \phi_\text{WF})
\end{align}
is the spin angular frequency component driven by the WF fields. Here $a_\text{WF}$ is a scaled WF field strength in units of the angular frequency. Finally, $\omega_\eta$ is the EDM-related term, namely
\begin{align}
	\omega_\eta (t) &= -\frac{q}{2m} \left( \frac{E_0}{c} - \beta B_0 \right) \eta(t) \equiv -\frac{d(t)}{S} E^*,
\end{align}
where $E^* \equiv E_0 - v B_0$ is the effective electric field, which is proportional to the EDM signal. With a highly relativistic beam $v \approx c$, a vertical magnetic field of 1 T provides an effective electric field of roughly $300$ MV/m by itself.

Given the reasonable assumptions $|\omega_\eta| \ll |a_\text{WF}|$ and $|\omega_\eta| \ll |\omega_{g-2}|$, we adopt the strategy used in the section \Rom{2} of Ref. \cite{Morse2013}. First we write down the exact solution for the radial and longitudinal spin components without the EDM, then plug those solutions into Eq. \eqref{eqn:Sy} to obtain the approximate time derivative of the vertical spin component. Without the EDM term, the radial and longitudinal spin components are given in forms:
\begin{align}
	S_x &= -\sin\left( \omega_{g-2}t + \frac{a_\text{WF}}{\omega_\text{WF}} (\sin(\omega_\text{WF}t + \phi_\text{WF}) - \sin\phi_\text{WF}) \right), \\
	S_s &= \cos\left( \omega_{g-2}t + \frac{a_\text{WF}}{\omega_\text{WF}} (\sin(\omega_\text{WF}t + \phi_\text{WF}) - \sin\phi_\text{WF}) \right),
\end{align}
where the initial polarization is assumed to be longitudinal $(S_s(0) = 1)$. Then plugging this $S_s$ into Eq. \eqref{eqn:Sy} yields
\begin{align}
	\dot{S_y} &\approx \frac{d(t)}{S} E^* \cos\left( \omega_{g-2}t + \frac{a_\text{WF}}{\omega_\text{WF}} \sin(\omega_\text{WF}t) \right)
\end{align}
by setting $\phi_\text{WF} = 0$ for a moment. This equation immediately shows the working principle of the frozen-spin method to measure the static EDM. When the WF is absent ($a_\text{WF} = 0$) and the spin is frozen ($\omega_{g-2} = 0$) then it leads to $\dot{S_y} \propto \eta_\text{DC}$, therefore, the vertical spin component accumulates with a slope proportional to the DC EDM. The argument is similar when the WF is present. We will use this equation separately for the DC and AC EDM terms to reveal the resonance conditions more explicitly. Recalling the EDM $\eta(t)$ in Eq. \eqref{eqn:eta}, it follows that
\begin{align} \label{eqn:S_y_DC}
	\left( \dot{S_y} \right)_\text{DC} &\approx \frac{d_\text{DC}}{S} E^* \cos\left( \omega_{g-2}t + \frac{a_\text{WF}}{\omega_\text{WF}} \sin(\omega_\text{WF}t) \right)
\end{align}
for the DC EDM term. One can see the resonance condition is $\omega_\text{WF} = \omega_{g-2}$, as studied in Ref. \cite{Morse2013}. To see it more manifestly, take $\omega_{g-2}t$ in the cosine bracket as a sawtooth wave with a period $2\pi/\omega_{g-2}$.  The remaining term of the argument is a sine function with a period $2\pi/\omega_\text{WF}$. If the periods of the two terms are not identical, then the phase argument uniformly sweeps from 0 to $2\pi$, which leads to $\langle \dot{S}_y \rangle = 0$ on average. The resonance happens when the two periods are identical, thus $\omega_\text{WF} = \omega_{g-2}$.

Similarly, the AC EDM term reads
\begin{widetext}
\begin{align}
	\left( \dot{S_y} \right)_\text{AC} &\approx \frac{d_\text{AC}}{S} E^* \cos(\omega_\text{axion} t) \cos\left( \omega_{g-2}t + \frac{a_\text{WF}}{\omega_\text{WF}} \sin(\omega_\text{WF}t) \right) \label{eqn:S_y_AC_1} \\
	&= \frac{d_\text{AC}}{2S} E^* \left[ \cos\left( (\omega_\text{axion} - \omega_{g-2})t - \frac{a_\text{WF}}{\omega_\text{WF}} \sin(\omega_\text{WF}t) \right) + \cos\left( (\omega_\text{axion} + \omega_{g-2})t + \frac{a_\text{WF}}{\omega_\text{WF}} \sin(\omega_\text{WF}t) \right) \right]. \label{eqn:S_y_AC}
\end{align}
\end{widetext}
Again we set $\phi_\text{axion} = 0$ for clarity. Restoring the WF phase and the axion phase in the above equations are straightforward, but of no importance at this moment. Using the same argument we made for the DC EDM term, we can clearly see that there are two resonance conditions: $\omega_\text{WF} = \omega_{g-2} \pm \omega_\text{axion}$. When the WF is absent, the resonance happens when $\omega_{g-2} = \omega_\text{axion}$ which is used in Ref. \cite{Chang2019}.

We point out that in the presence of the WF, the average slope of the vertical spin component $\left\langle \dot{S}_y \right\rangle$ in the resonance condition is multiplied by the following factor:
\begin{align} \label{eq:C_WF}
	C_\text{WF} \equiv \left\langle \cos\left( \omega t + \frac{a}{\omega} \sin(\omega t) \right) \right\rangle,
\end{align}
where we dropped the subscript WF in the right hand side for a moment. Interestingly, it shows that $C_\text{WF} = -J_1 (a/\omega)$ where $J_1$ is the Bessel function of the first kind. See App. \ref{app:CWF_derivation} for the derivation. The maximum absolute value of $C_\text{WF}$ is therefore roughly 0.59 when $a/\omega \approx 1.84$. Eventually, we have the following expression for the vertical spin slope (EDM signal) on resonance.
\begin{align} \label{eq:EDMsignal}
	\omega_d = -\frac{d_\text{AC}}{2S} E^* J_1 \left( \frac{a_\text{WF}}{\omega_\text{WF}} \right)
\end{align}

\begin{table*}[t]
	\centering
	\caption{Various methods seeking to probe either the stationary (DC) or oscillating (AC) EDM in storage rings. All methods exploit the vertical spin resonance for a specific resonance condition. $s$ is the spin quantum number. `srEDM' stands for the storage ring EDM experiments.}
	\label{tab:various_methods}
	\begin{tabular}{lcccc}
		\hline\hline
		Method & srEDM & srEDM + WF & srAxionEDM & srAxionEDM + WF \\
		\hline
		Measurement target & $d_\text{DC}$ & $d_\text{DC}$ & $d_\text{AC}$ & $d_\text{AC}$ \\
		Resonance condition & $\omega_{g-2} = 0$ & $\omega_{g-2} = \omega_\text{WF}$ & $\omega_{g-2} = \omega_\text{axion}$ & $\omega_{g-2} = | \omega_\text{axion} \pm \omega_\text{WF} |$ \vspace{4pt} \\ \vspace{4pt}
		Spin vertical slope ($\omega_d$) & $\dfrac{d_\text{DC}}{s \hbar} E^*$  & $\dfrac{d_\text{DC}}{s \hbar} E^* C_\text{WF}$ & $\dfrac{d_\text{AC}}{2 s \hbar} E^*$ & $\dfrac{d_\text{AC}}{2 s \hbar} E^* C_\text{WF}$ \\
		References & \cite{Anastassopoulos2016} & \cite{Morse2013} & \cite{Chang2019, Pretz2020} & This work \\
		\hline\hline
	\end{tabular}
\end{table*}

An overview of the various methods used to measure EDM in storage rings, including this work, is provided in Table \ref{tab:various_methods}.

\section{Spin Tracking Simulation}
In this section a spin tracking simulation for the reference particle was performed as a proof-of-principle. A deuteron, whose magnetic anomaly is $G_d = -0.143$, was simulated in a purely magnetic ring with a radius of 30 m. The reference momentum was given to be 1 GeV/c. The corresponding $g-2$ frequency is around 121 kHz. Also, the axion frequency was arbitrarily assumed to be 180 kHz, which corresponds to an axion mass of 0.7 neV/c$^2$. The initial setup parameters are summarized in Table \ref{tab:initial_params}.

\begin{table}[t]
	\centering
	\caption{The initial parameters for the spin tracking simulation with the RF Wien Filter.}
	\label{tab:initial_params}
	\begin{tabular}{lr}
		\hline\hline
		Parameter\phantom{blahblahblahblah} & Type or Value \\
		\hline
		Particle & Deuteron \\
		Magnetic anomaly ($G$) & $-0.143$ \\
		Ring radius & 30 m \\
		Magnetic field strength & 0.111 T \\
		Reference momentum & 1 GeV/c \\
		Reference velocity ($\beta$) & 0.47 \\
		Reference Lorentz factor ($\gamma$) & 1.133 \\
		$g-2$ frequency & 121 kHz \\
		Axion frequency & 180 kHz \\
		\hline\hline
	\end{tabular}
\end{table}

\begin{figure}[t]
	\centering
	\includegraphics[width=0.45\textwidth]{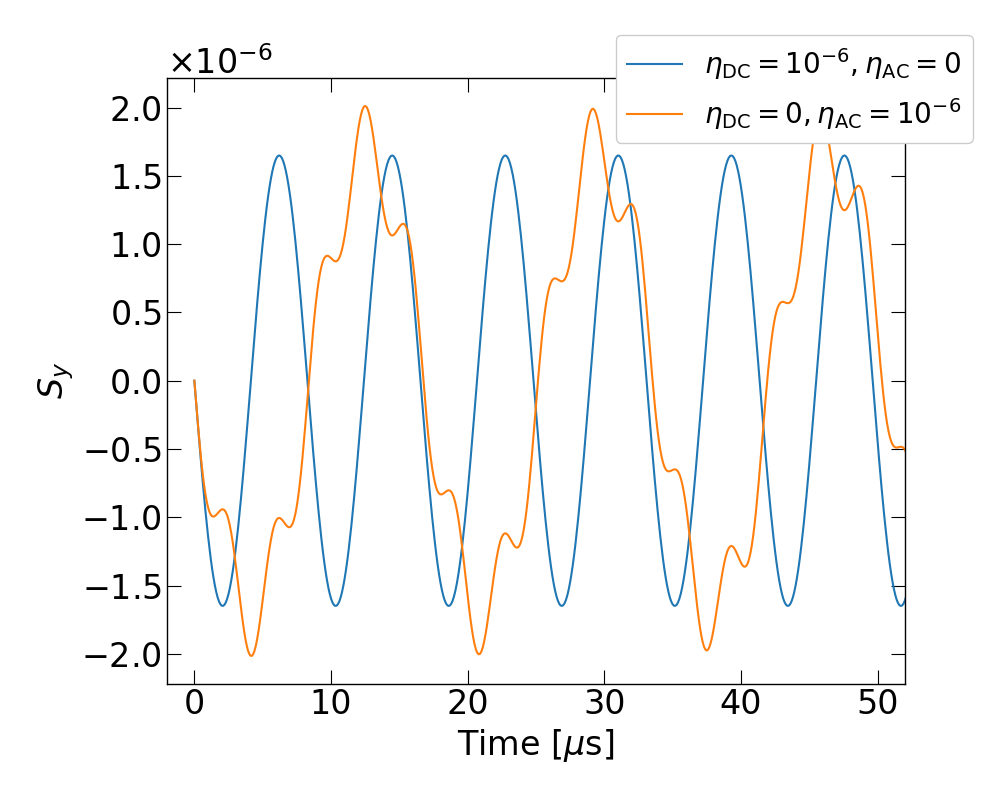}
	\caption{Vertical spin component versus time, with only the DC EDM of $\eta_\text{DC} = 10^{-6}$ (blue), and only the AC EDM of $\eta_\text{AC} = 10^{-6}$ (orange).}
	\label{fig:Sy_noWF}
\end{figure}

\begin{figure}[t]
	\centering
	\includegraphics[width=0.45\textwidth]{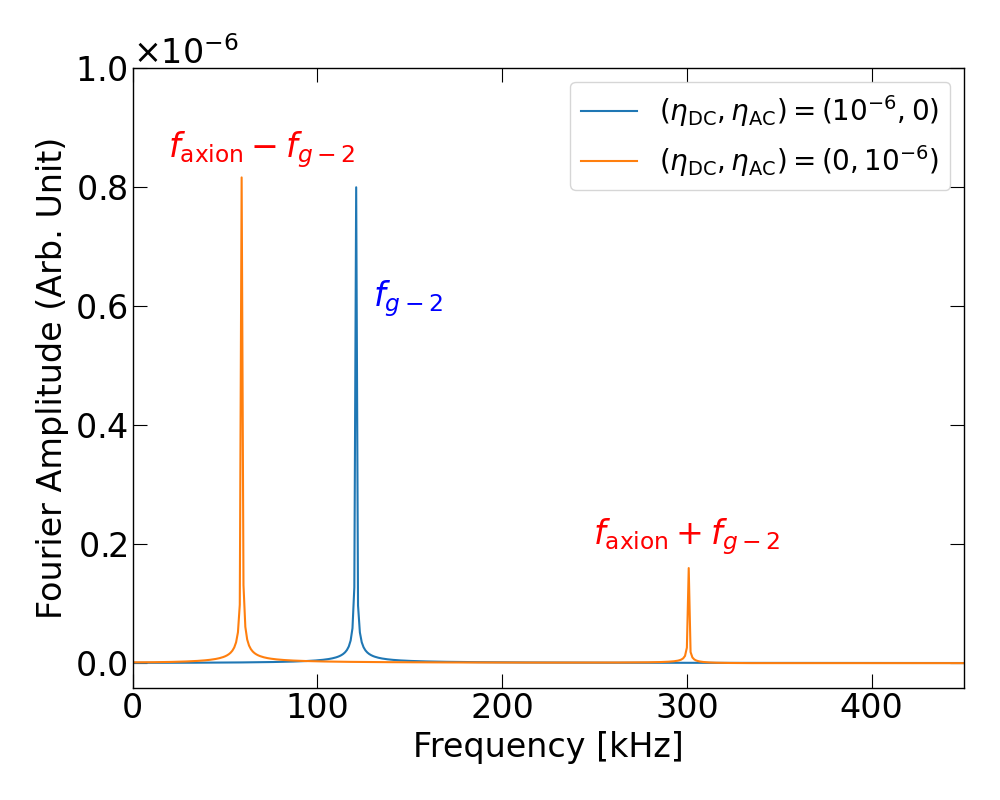}
	\caption{Fourier spectra of the vertical spin component, with only the DC EDM of $\eta_\text{DC} = 10^{-6}$ (blue) and only the AC EDM of $\eta_\text{AC} = 10^{-6}$ (orange). In the simulation the blue peak is located at $f_{g-2} \approx 120$ kHz and the orange peaks are located at $f_{g-2} \pm f_\text{axion}$ where $f_\text{axion} = 180$ kHz.}
	\label{fig:SyFFT_noWF}
\end{figure}

First, the simulation was conducted without applying the WF. Fig. \ref{fig:Sy_noWF} shows the vertical spin component ($S_y$) of the reference particle in the presence of the EDM, with blue when there is only the DC component of the EDM (with $\eta_\text{DC} = 10^{-6}$) and orange for the AC component only ($\eta_\text{AC} = 10^{-6}$). Fig. \ref{fig:SyFFT_noWF} shows their Fourier spectra. One can immediately see $\eta_\text{DC}$ is responsible for the $g-2$ frequency and $\eta_\text{AC}$ is for its sidebands with the axion frequency: $f_\text{axion} \pm f_{g-2}$. The peak at the higher frequency sideband, $f_\text{axion} + f_{g-2} \approx 300$ kHz, clearly looks much smaller than the other sideband, $f_\text{axion} - f_{g-2} \approx 60$ kHz. The reason is quite straightforward. As implied in Eq. \eqref{eqn:S_y_AC_1}, without the WF ($a_\text{WF} = 0$), one gets the time derivative of the vertical spin component asymptotically proportional to $\cos(\omega_\text{axion}t) \cos(\omega_{g-2} t)$. Integrating this expression, one obtains two sidebands whose amplitudes are divided by each frequency. In this case, the ratio between the higher and lower sidebands is roughly 5, which is the magnitude ratio of peaks for the two sidebands in Fig. \ref{fig:SyFFT_noWF}.

\begin{figure}[t]
	\centering
	\includegraphics[width=0.45\textwidth]{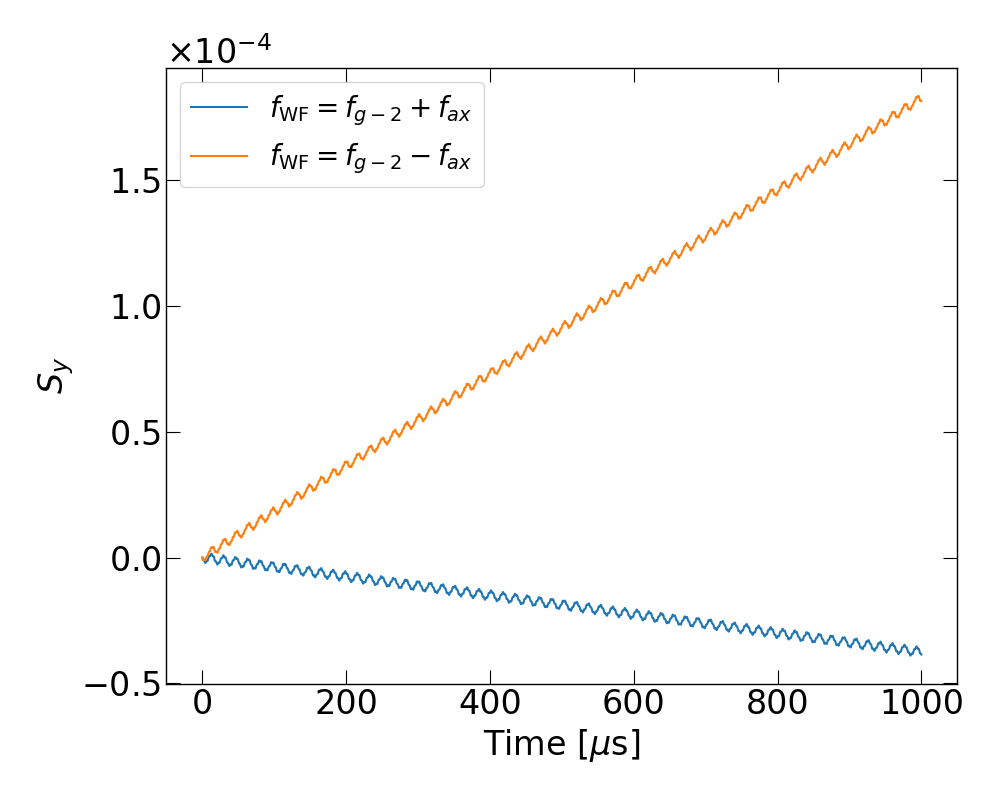}
	\caption{Vertical spin component versus time, when an RF Wien Filter of 1 MV/m electric field strength was applied at the higher sideband $f_\text{axion} + f_{g-2}$ (blue) and lower sideband $f_\text{axion} - f_{g-2}$ (orange). The Wien Filter is continuously located on the storage ring in the simulation.}
	\label{fig:Sy_WF}
\end{figure}

Next, we want to observe the vertical spin resonance in the presence of the WF operating at one of the sidebands of the axion and $g-2$ frequency. The WF was assumed to occupy all parts of the storage ring continuously, and its electric field strength was set to 1 MV/m. As Fig. \ref{fig:Sy_WF} represents, the spin vertical component accumulates in one direction when the WF is applied at either one of the two sidebands. The directions of the accumulation of the spin vertical component for two cases are opposite; they depend on the sign of $\omega_\text{axion} \pm \omega_{g-2}$ and $\omega_\text{WF}$. And the slope depends on the sideband frequency like the previous argument as well: $\langle \dot{S}_y \rangle \propto 1/\omega_\text{WF}$.

\begin{figure}[t]
	\centering
	\includegraphics[width=0.45\textwidth]{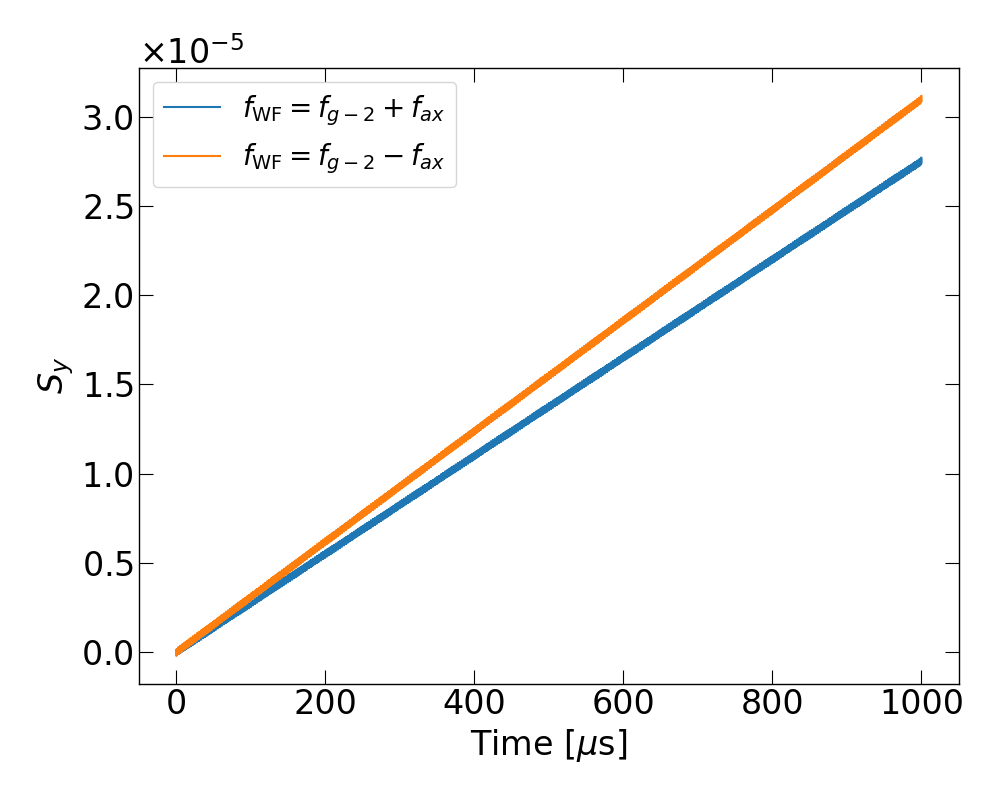}
	\caption{Vertical spin component versus time for a proton with a momentum 1 GeV/c, when an RF Wien Filter of 1 MV/m electric field strength was applied at the higher sideband $f_\text{axion} + f_{g-2}$ (blue) and lower sideband $f_\text{axion} - f_{g-2}$ (orange). The Wien Filter is continuously located on the storage ring in the simulation.}
	\label{fig:Sy_WF_proton}
\end{figure}

So far we have used a deuteron for the simulation. It is worth trying a proton as well, which has a magnetic anomaly $G_p = 1.793$. Its mass is around 938 MeV/c$^2$, almost half that of the deuteron. Assuming the same reference momentum $p = 1$ GeV/c and a WF electric field strength of 1 MV/m, the corresponding $g-2$ frequency for the proton becomes 3 MHz, which is 25 times larger than that of the deuteron. As a result, the vertical spin component growth rate for the proton is around an order of magnitude smaller than that of the deuteron, as shown in Fig. \ref{fig:Sy_WF_proton}.

\begin{figure}[t]
	\centering
	\includegraphics[width=0.45\textwidth]{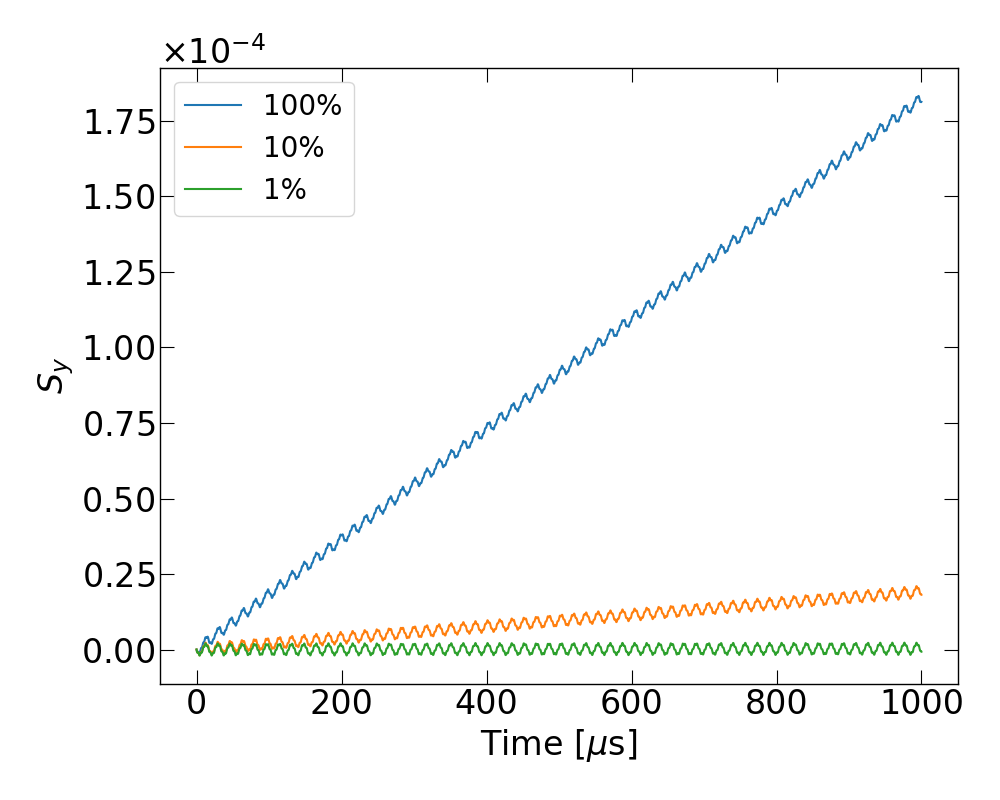}
	\caption{Vertical spin component versus time for different Wien Filter azimuthal occupancy: continuously located along the storage ring (blue), taking only 10\% of the ring azimuth (orange) and 1\% (green).}
	\label{fig:Sy_WF_WFtheta}
\end{figure}

We also have briefly tested the WF taking only a fraction of the storage ring azimuth, instead of continuously located along the ring. Figure \ref{fig:Sy_WF_WFtheta} shows the vertical spin component on resonance with different WF azimuthal occupancy, which is consistent with the natural expectation; the slope scales the same with the WF occupancy.

\section{Systematic Error Studies}

	\subsection{Wien Filter Misalignment}
\begin{figure}[t]
	\centering
	\includegraphics[width=0.45\textwidth]{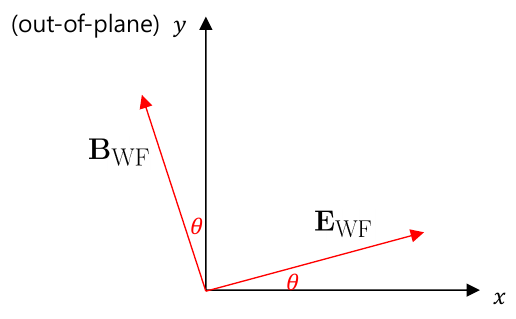}
	\caption{The electromagnetic field from the RF Wien Filter, tilted by an angle $\theta$. Ideally, the electric field of the Wien Filter designed for this storage ring should only have a radial component ($\hat{e}_x$) and the magnetic field should only have a vertical component ($\hat{e}_y$). But this small angle $\theta$ might be there because of the misalignment.}
	\label{fig:tilted_WF}
\end{figure}

\begin{figure}[t]
	\centering
	\includegraphics[width=0.45\textwidth]{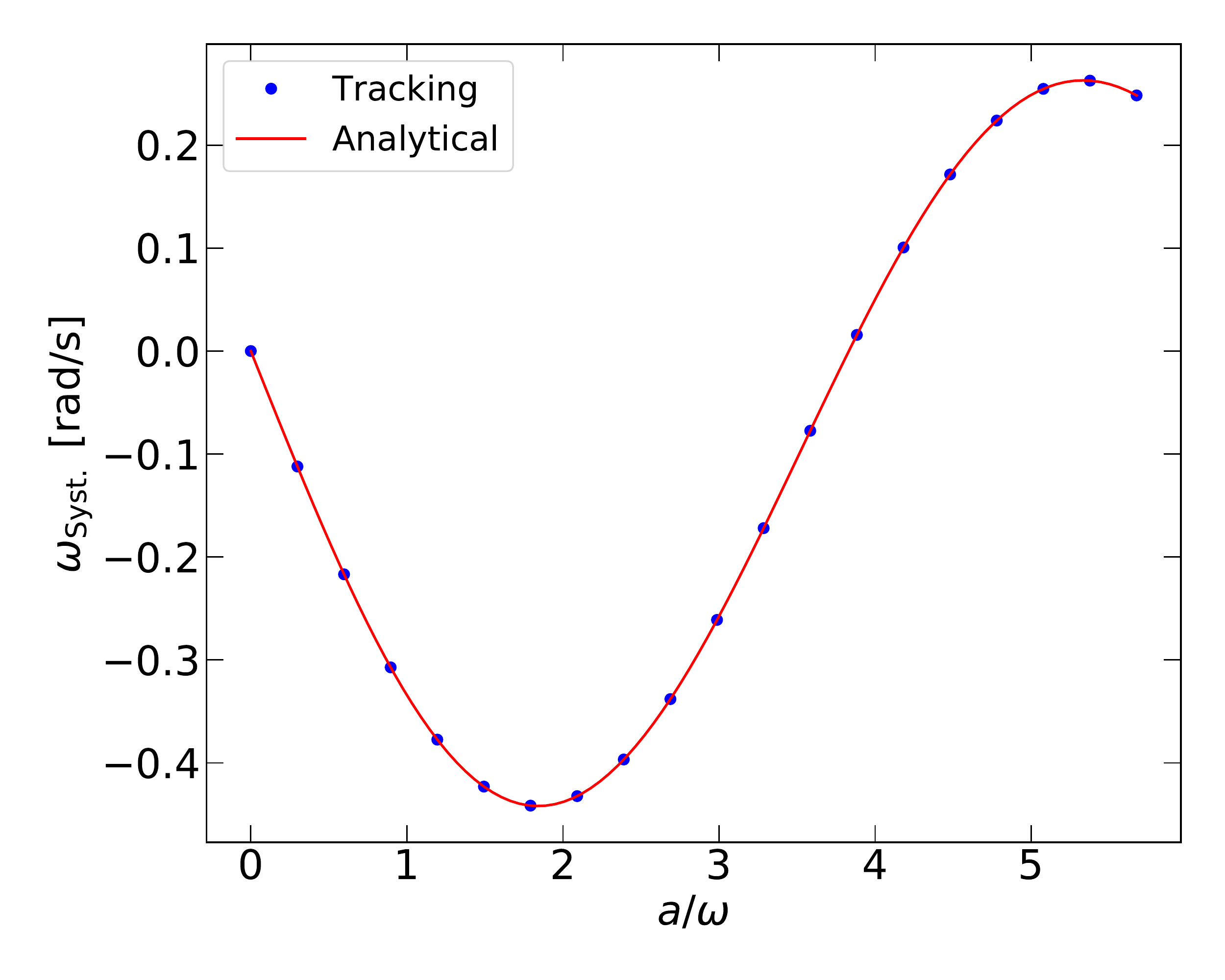}
	\caption{The vertical spin slope driven by the WF misalignment, obtained from the spin tracking simulation (blue circles) and by the analytical expression in Eq. \eqref{eq:C_Syst} (red curve). Both the DC and AC EDM were set to 0 for the spin tracking. The misalignment angle $\theta_m$ is 1 $\mu$rad.}
	\label{fig:omega_syst}
\end{figure}

In Eqs. \eqref{eqn:S_y_DC} and \eqref{eqn:S_y_AC}, we found that the DC component of the EDM is sensitive to the $g-2$ frequency of WF, and the AC component is sensitive to the sidebands of the $g-2$ and axion frequency, respectively. It turns out that the former case is vulnerable to a systematic error from WF misalignment\cite{Morse2013}. When the RF electric and magnetic fields point in a slightly tilted direction, as illustrated in Fig. \ref{fig:tilted_WF}, the resulting radial component of the RF magnetic field can drive the resonance of the vertical spin component, mimicking the EDM signal. Specifically, the radial component of the spin angular frequency induced by WF misalignment reads
\begin{align}
	\omega_{sx}^\text{WF} &= -\frac{q}{m} \left[ \left( G + \frac{1}{\gamma} \right) B_x^\text{WF} + \left( G + \frac{1}{\gamma + 1} \right) \frac{\beta E_y^\text{WF}}{c} \right] \\
	&= \theta_m a_\text{WF} \cos(\omega_\text{WF} t + \phi_\text{WF}),
\end{align}
where $\theta_m$ is the tilted angle of the WF. This leads to the slope of the vertical spin component coming from the WF misalignment systematic effect,
\begin{align}
	\left( \dot{S_y} \right)_\text{Syst.} = -\theta_m a_\text{WF} \cos(\omega_\text{WF} t + \phi_\text{WF}) S_s.
\end{align}
In the resonance condition $\omega_\text{WF} = \omega_{g-2}$, it leads to a systematic vertical spin slope proportional to
\begin{align} \label{eq:C_Syst}
	C_\text{Syst.} &\equiv \left\langle \cos(\omega t) \cos\left( \omega t + \frac{a}{\omega} \sin(\omega t) \right) \right\rangle
\end{align}
having additional $\cos(\omega t)$ compared to Eq. \eqref{eq:C_WF}. It is also shown that $C_\text{Syst.}$ can be represented by the Bessel function of the first kind: $C_\text{Syst.} = \frac{1}{2} \left\{ J_0 \left( \frac{a}{\omega} \right) + J_2 \left( \frac{a}{\omega} \right) \right\}$. The derivation is provided in App. \ref{app:CWF_derivation}. Therefore, we have a systematic vertical spin slope, as follows.
\begin{align} \label{eq:omega_syst}
	\left\langle \left( \dot{S_y} \right)_\text{Syst.}  \right\rangle = -\frac{1}{2} \theta_m a \left[ J_0 \left( \frac{a}{\omega} \right) + J_2 \left( \frac{a}{\omega} \right) \right].
\end{align}
Equation \eqref{eq:omega_syst} was confirmed by the spin tracking simulation, as shown in Fig. \ref{fig:omega_syst}. The basic setup was the same as in Table \ref{tab:initial_params}, except there was a WF misalignment angle of 1 $\mu$rad for this case. Both the DC and AC EDM were set to 0 to make sure the effect was purely from the systematics. The slopes obtained from the spin tracking simulation were in an excellent agreement with the analytical calculation. Typically, this systematic effect can be frustrating if we cannot control the WF misalignment angle to really small tolerances, on the order of nanoradians, but the present method is largely free from it.

\begin{figure}[htb]
	\centering
	\includegraphics[width=0.45\textwidth]{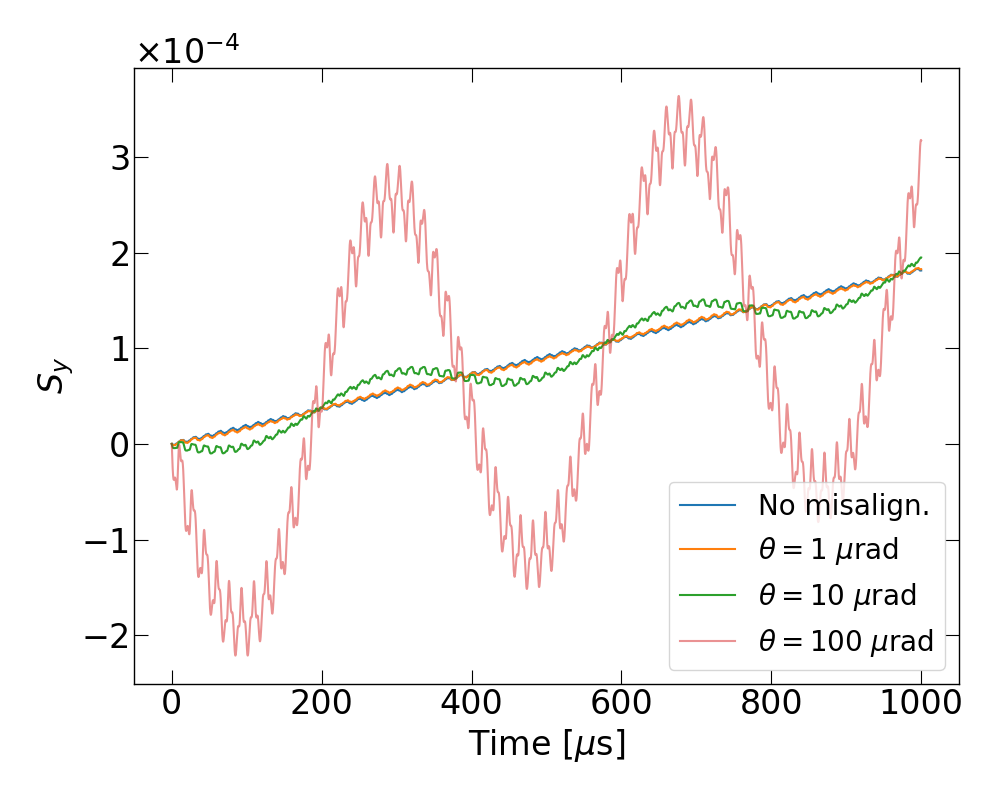}
	\caption{The vertical spin component versus time, when there is only the AC EDM component ($\eta_\text{AC} = 10^{-6}$) and the RF Wien Filter frequency is the lower sideband $f_\text{axion} - f_{g-2}$. The Wien Filter is assumed to be perfectly aligned (blue), misaligned by 1 $\mu$rad (orange), 10 $\mu$rad (green) and 100 $\mu$rad (red). The tilted direction is the same in all cases, following Fig. \ref{fig:tilted_WF}.}
	\label{fig:Sy_misalignedWF}
\end{figure}

This systematic effect can be avoided if the WF frequency is not close to the $g-2$ frequency, at $\omega_\text{axion} \pm \omega_{g-2}$, as confirmed in Fig. \ref{fig:Sy_misalignedWF}. The WF with the lower sideband frequency was applied in the presence of just the AC EDM. Each color represents the degree of the misalignment, from 0 to 100 $\mu$rad. No matter what the tilted angle $\theta$, the vertical spin component grows with the same average slope. The large fluctuations on the growth for the case of 100 $\mu$rad, for example, are not a problem since the average slope is what matters for detecting the vertical spin resonance.

When the misalignment angle for the electric field is different from that of the magnetic field generated by the WF, the particles experience the Lorentz force, and this can be immediately corrected through precise beam position monitoring.

	\subsection{Intrinsic Resonances and Field Errors}
\begin{figure}[htb]
	\centering
	\includegraphics[width=0.45\textwidth]{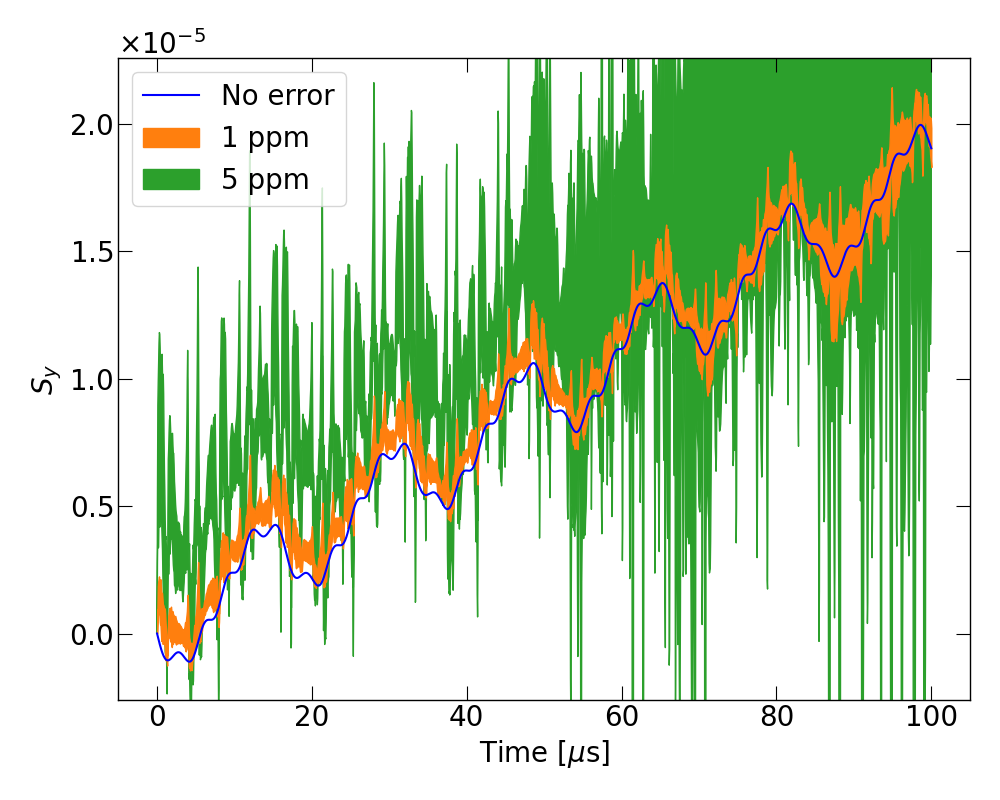}
	\caption{The fluctuations of the vertical spin component versus time in presence of the random field errors. The blue curve is the reference plot without the field errors. The magnetic field errors up to the octupole ($k=4$), 8-th azimuthal harmonics ($N=8$) were implemented, where the relative strength $b_{k, N}/B0$ are chosen to be 1 (orange) and 5 (green) parts-per-million (ppm), respectively. 10 independent series of simulation were conducted for each category, and the fluctuation between the maximum and the minimum values of the vertical spin component at each time bin is plotted.}
	\label{fig:Sy_random_field_error}
\end{figure}

There are also intrinsic systematic error sources in particle accelerators, namely the betatron tune and spin resonances. A general condition of the spin resonance is given as\cite{Mane2005, Conte2008}
\begin{align}
	N_\text{spin} \nu_\text{spin} + N_x \nu_x + N_y \nu_y + N_\text{sync} \nu_\text{sync} = N,
\end{align}
where $\nu_\text{spin}, \nu_x, \nu_y, \nu_\text{sync}$ are the spin, horizontal, vertical and synchrotron tunes, respectively, and the coefficients and $N$ are integers. Although not all set of the tunes satisfying the above condition lead to the resonance strong enough to depolarize the beam, it is recommended to avoid the resonances as best as one can, especially for the low-order ones with relatively small $|N|$s. This in general can be done by adjusting the focusing field index and carefully setting the spin precession frequency.

The field errors are closely related to the intrinsic resonances as well, since the beam experiences them periodically. The field error can be represented using the multipole and Fourier expansion,
\begin{align}
	B_x (x, y, s) &= \sum_{k=1, N=1} b_{k, N} \,\Im \left( \frac{x + iy}{r_a} \right)^{k-1} \cos \left( N \frac{s}{R} + \phi_N \right), \\
	B_y (x, y, s) &= \sum_{k=1, N=1} b_{k, N} \,\Re \left( \frac{x + iy}{r_a} \right)^{k-1} \cos \left( N \frac{s}{R} + \phi_N \right),
\end{align}
which describes the normal $2k$-pole, $N$-th azimuthal harmonic magnetic field, where $r_a$ is the beam storage acceptance radius and $R$ is the storage ring radius. One can swap the real and imaginary parts and put additional negative sign to $B_y$ to obtain the skewed multipole components.

We studied the effect of the field errors by implementing the above field components with randomly generated $b_{k, N}$ up to octupole ($k=4$), 8-th azimuthal harmonics ($N=8$), where $r_a$ is chosen to be 10 cm. Although the relative strength $b_{k, N}/B_0$ typically decreases as the order gets higher, we set the same scales for all $k$ and $N$s to avoid complications. The result is shown in Fig. \ref{fig:Sy_random_field_error}, where the blue curve is the nominal $S_y$ on resonance without the field error as reference. The other curves represent the maximum value of the random-generated relative field strength $b_{k, N}/B_0$, which are 1 parts-per-million (ppm) for the orange and 5 ppm for the green, respectively. The two curves with the field errors are actually plotting the fluctuations, filling between the maximum and minimum values of $S_y$ in 10 independent series of simulation done for each of them. It is clearly shown that the spin motion is fairly stable with the field errors of order of 1 ppm, but shows larger fluctuations with 5 ppm. These fluctuations are averaged out with long storage times.

In general, if the proposed experiment encounters a systematic spin resonance due to many potential sources such as intrinsic tune resonance or Berry’s phase, one can figure out whether it is a systematic noise or a real signal by readjusting the $g-2$ frequency and the Wien Filter frequency targeting the same axion frequency.

\section{Estimation of Statistical Sensitivity} \label{sec:sensitivity}

\begin{figure}[t]
	\centering
	\includegraphics[width=0.5\textwidth]{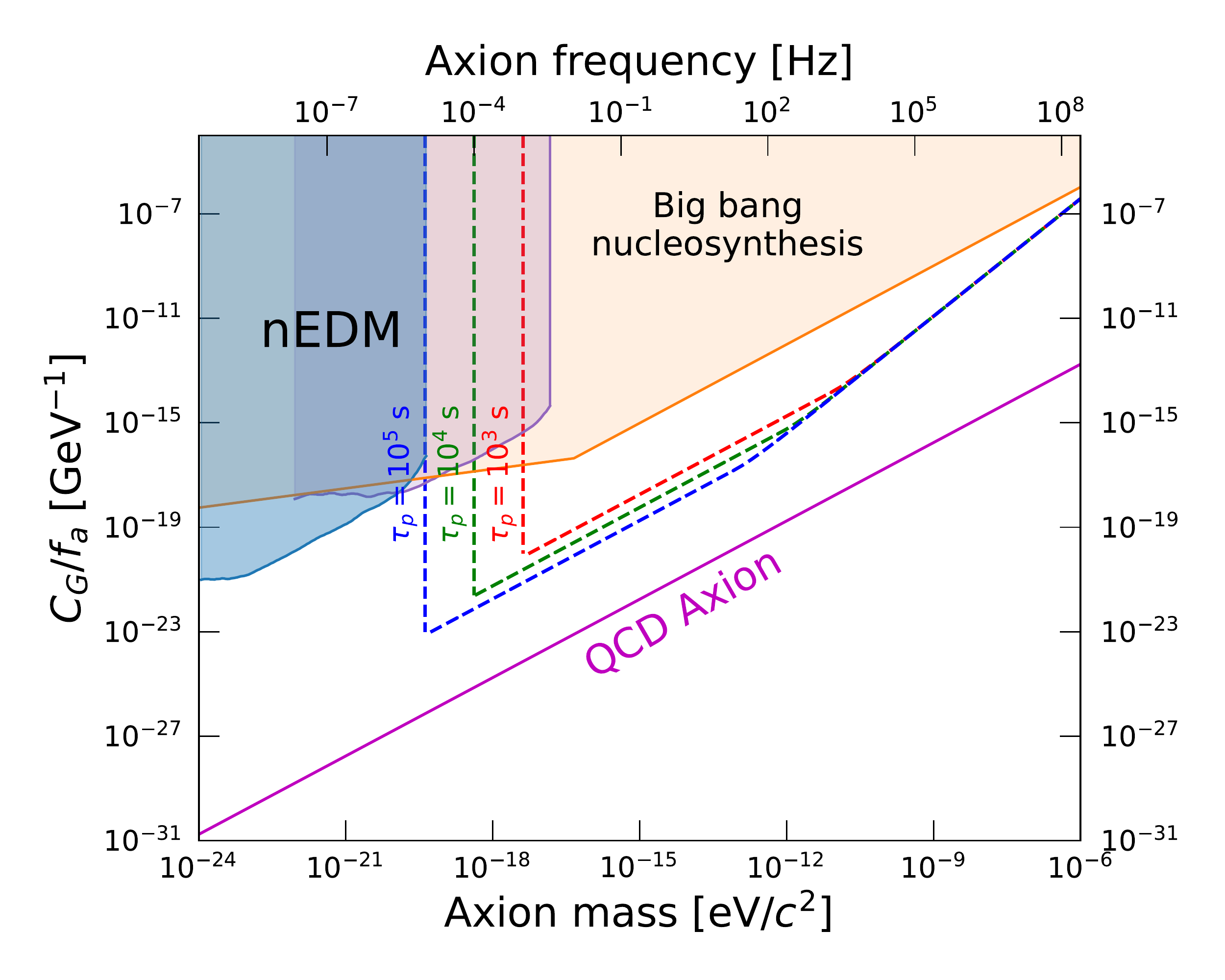}
	\caption{The projected sensitivity to the axion-gluon coupling strength in the axion parameter space for axionlike dark matter. The colored dashed lines indicate the present study, depending on the spin coherence time $\tau_p$: $10^3$ s (red), $10^4$ s (green) and $10^5$ s (blue). It assumes the given values in Eq. \eqref{eq:sigma_d_nominal_value} for the proton. The integrated measurement time at each frequency is one year. The filled regions that were already excluded by the neutron EDM (nEDM) experiment\cite{Abel2017} and the big bang nucleosynthesis\cite{Blum2014} are shown as references, as well as the QCD axion band in Eq. \eqref{eq:QCDaxion}.}
	\label{fig:Sensitivity}
\end{figure}

The statistical sensitivity of the oscillating EDM has been derived in a rigorous manner in App. \ref{app:Sensitivity}.
\begin{align} \label{eq:sigma_d_sensitivity}
	\sigma_d = \frac{4.67 s \hbar}{P_0 A E^* C_\text{WF} \sqrt{\kappa N_\text{cyc} T_\text{exp} \tau_p}},
\end{align}
where $P_0$ is the initial beam polarization, $A$ is the analyzing power, $E^*$ is the equivalent electric field in the storage ring, $C_\text{WF}$ is the coefficient determined by the WF performance, $\kappa$ is the polarimeter efficiency, $N_\text{cyc}$ is the number of stored particles in one cycle (single measurement), $T_\text{exp}$ is the total experimental period and $\tau_p$ is the spin coherence time. Plugging in the typical experimental numbers in the ideal situation, one obtains
\begin{widetext}
\begin{align} \label{eq:sigma_d_nominal_value}
	\sigma_d = 9.3 \times 10^{-31} \text{ [} e \cdot \text{cm]}
	\left( \frac{s}{1/2} \right) \left( \frac{0.8}{P_0} \right) \left( \frac{0.6}{A} \right) \left( \frac{100 \text{ MV/m}}{E^*} \right) \left( \frac{0.59}{C_\text{WF}} \right) \sqrt{\left( \frac{1.1\%}{\kappa} \right) \left( \frac{10^{11}}{N_\text{cyc}} \right) \left( \frac{1 \text{ yr}}{T_\text{exp}} \right) \left( \frac{10^3 \text{ s}}{\tau_p} \right)}
\end{align}
\end{widetext}
for the case of the proton. This calculation is done assuming we can obtain a large $C_\text{WF}$ for all targeting frequencies in the axion parameter space. For the realistic case, $C_\text{WF}$ is more restrictive depending on the WF performance. For instance, $a_\text{WF}$ should be proportional to the azimuthal fraction the WF occupies in the storage ring. But we do not cover the technical difficulties and details to achieve the maximum $C_\text{WF}$ in this paper.

In a search for axionlike dark matter, we can exclude the axion-gluon coupling parameter space with a sensitivity proportional to $\sigma_d$. The QCD axion has the relationship\cite{diCortona2016}
\begin{align} \label{eq:QCDaxion}
	m_a^\text{QCD} \approx 5.7 \text{ $\mu$eV } \left( \frac{10^{12} \text{ GeV}}{(f_a/C_G)^\text{QCD}} \right)
\end{align}
where $C_G$ is a model-dependent dimensionless coefficient of the axion-gluon coupling Lagrangian and $f_a$ is the symmetry breakdown scale. Exploiting this relation, the sensitivity to $(C_G/f_a)$ that we can exclude from the parameter space if the EDM is not discovered with the uncertainty $\sigma_d$, is given as
\begin{align} \label{eq:CGfa}
	\left( \frac{C_G}{f_a} \right)_\text{exc.} = \left( \frac{C_G}{f_a} \right)^\text{QCD} \frac{\sigma_d}{\left| d_n^\text{QCD} \right|}
\end{align}
where $d_n^\text{QCD} \approx 9 \times 10^{-35} \cos(m_a t)$ [$e \cdot$ cm] holds for the QCD axion\cite{Graham2013}.

The projected sensitivity on $C_G/f_a$ is shown in Fig. \ref{fig:Sensitivity}, indicated by the colored dashed lines. These were obtained from Eq. \eqref{eq:CGfa} using Eq. \eqref{eq:sigma_d_nominal_value} for the given values within, except the spin coherence time. The minimum axion frequency that one can scan is determined by a single measurement time, as long as it is free from systematic effects. Therefore, a higher spin coherence time not only improves the sensitivity at a given frequency but also widens the scanning area. The optimum single measurement time to minimize the statistical uncertainty of the repeated measurement is shown to be $\tau_p/2$, as derived in App. \ref{app:Sensitivity}. There are kinks, after which the projected sensitivity starts to decrease fast. This is because the single measurement time $T$ has to be smaller than $\tau_p/2$ after this point, as the axion phase decoheres faster than $\tau_p/2$. Explicitly, $T$ is given by
\begin{align}
	T = \min \left\{ \frac{Q_\text{axion}}{f_\text{axion}}, \frac{\tau_p}{2} \right\},
\end{align}
where $Q_\text{axion}$ is the axion quality factor, which was assumed to be $10^{6}$\cite{Krauss1985, Turner1990}.

\begin{figure}[t]
	\centering
	\includegraphics[width=0.5\textwidth]{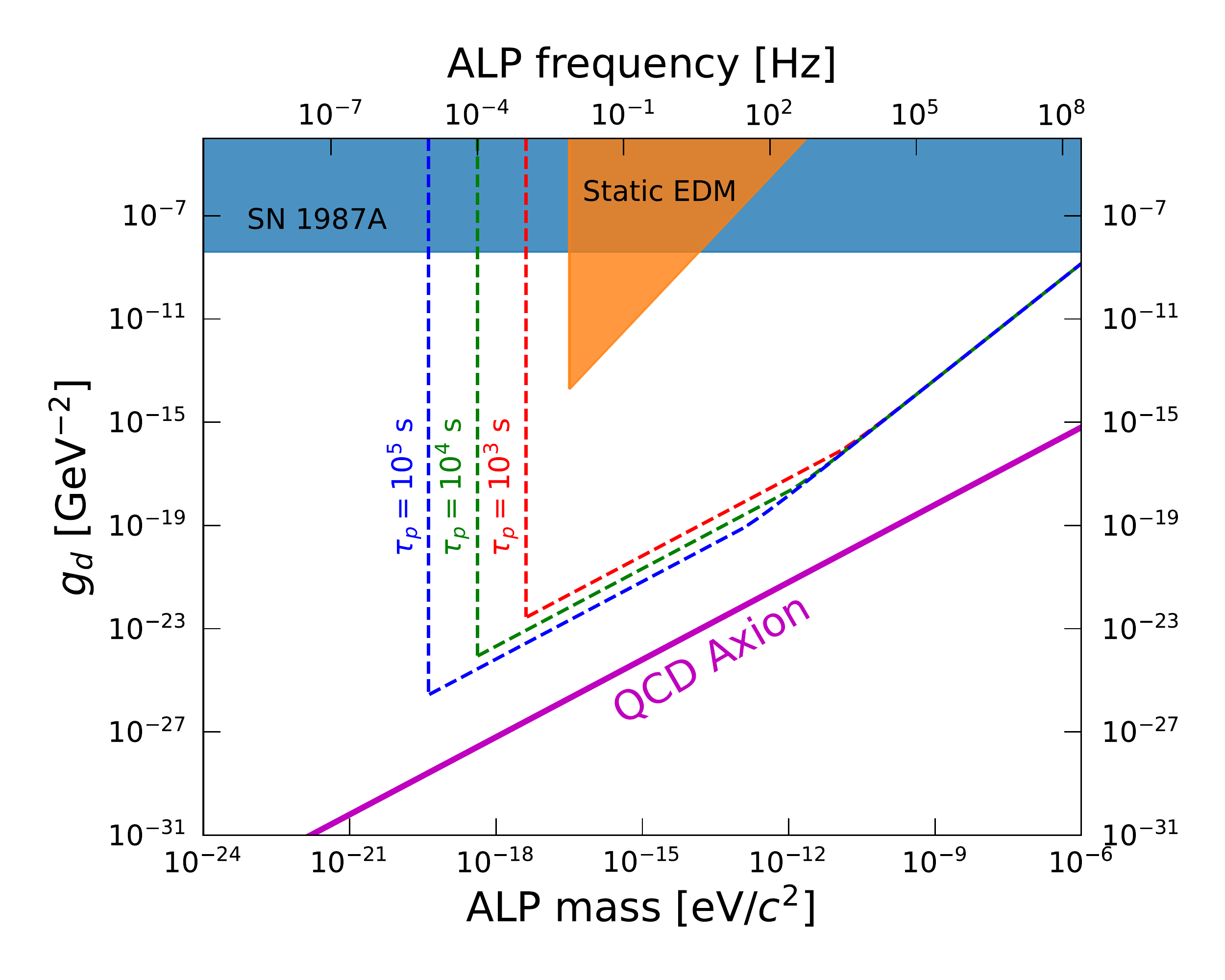}
	\caption{The projected sensitivity to the ALP-nucleon EDM coupling strength ($g_d$) in the ALP parameter space for axionlike dark matter. The colored dashed lines indicate the present study, depending on the spin coherence time $\tau_p$: $10^3$ s (red), $10^4$ s (green) and $10^5$ s (blue). It assumes the given values in Eq. \eqref{eq:sigma_d_nominal_value} for the proton. The integrated measurement time at each frequency is one year. The filled regions that were excluded by excess cooling in SN1987A (light blue) and the static EDM measurement (orange) are adapted from Ref. \citep{Graham2013}, with proper extension to the range in the present parameter space.}
	\label{fig:Sensitivity_gd}
\end{figure}

On the other hand, the parameter space for the new coupling between the axion-like particles (ALPs) and the nucleon, $g_d$, can be scanned. This coupling, which is directly responsible for the oscillating nucleon EDM, appears in the Lagrangian\citep{Graham2013}
\begin{align}
	\mathcal{L} \ni -\frac{i}{2} g_d a \bar{N} \sigma_{\mu\nu} \gamma_5 N F^{\mu\nu},
\end{align}
where $a$ is the ALP field interacting with the nucleon $N$. Assuming the ALP makes up all of the local dark matter with a density $\rho_\text{DM} \approx 0.3 \text{ GeV/cm$^3$}$, the nucleon EDM is given by\citep{Graham2013}
\begin{align}
	d_n \approx \left( 1.4 \times 10^{-25} \; e \cdot \text{cm} \right) \left( \frac{\text{eV}}{m_a} \right) \left( \frac{g_d}{\text{GeV}^{-2}} \right) \cos(m_a t).
\end{align}
Figure \ref{fig:Sensitivity_gd} shows the projected sensitivity to the EDM coupling $g_d$. The constraints from the excess cooling in SN198A and the static EDM measurement are provided in Ref. \citep{Graham2013}, where we have set the lower bound of the constraint from the static EDM to 1/130 Hz as the experimental time for a single shot was $t_\text{shot} = 130$ seconds\cite{Baker2006}.

Recently, there was a study offering new predictions of the ALPs coupling, called ALPs cogenesis\cite{Co2021}. Its projected region in the parameter space was above the QCD axion band, strengthening the motivation for scanning the axion parameter space above the QCD axion band.

\section{Conclusion}
Employing an RF Wien Filter in the storage ring EDM method provides a powerful method for probing an oscillating EDM. We revealed that vertical spin resonance happens when the Wien Filter is operating at the frequency of the sidebands of the axion and the $g-2$ frequency, $f_\text{axion} \pm f_{g-2}$. The approximated analytic solution for the spin resonance agreed well with the simulation results. Scanning the axion frequency would be straightforward, easily performed by tuning the Wien Filter frequency. This method avoids the large systematic effect that arises when the Wien Filter frequency is close to the $g-2$ frequency, as confirmed by both analytical calculations and spin tracking simulations. Even though when the Wien Filter frequency is close to the $g-2$ frequency to search for low-mass axion ($f_\text{axion} < 1$ Hz), the systematic effect from Wien Filter misalignment is well understood and can be corrected precisely.

A systematic effect from random field errors was also studied, which did not show a critical influence on the vertical spin component at least up to 5 ppm. Nonetheless, further intensive numerical studies are necessary under more realistic lattice and beam conditions to understand the details of all systematic effects before conducting an experiment.

This particular idea of introducing an RF Wien Filter to look for the axion-induced oscillating EDM might be of interest for frozen-spin proton and deuteron EDM experiments in storage rings, because it allows physics probing data to be taken simultaneously from the static DC EDM and the oscillating AC EDM. This method can be applied to existing storage rings, e.g., the muon $g-2$ experiment at Fermilab\cite{Abi2021}, by storing polarized proton or deuteron beams.

\begin{acknowledgements}
	This work was supported by IBS-R017-D1-2021-a00. We appreciate informative discussions with members of the storage ring EDM collaboration.
\end{acknowledgements}

\onecolumngrid
\appendix
\section{Derivation of $C_\text{WF}$ and $C_\text{Syst.}$} \label{app:CWF_derivation}
The Bessel function of the first kind has the following integral representation\cite{Temme1996}:
\begin{align}
	J_n(x) = \frac{1}{2\pi} \int_0^{2\pi} \cos(n\theta - x \sin\theta) \; \dd \theta, \qquad n=0, 1, 2 \cdots.
\end{align}
Substituting $\theta = \omega t$ and $T = 2\pi/\omega$, one obtains
\begin{align}
	J_n(x) = \frac{1}{T} \int_0^T \cos\bigg( n \omega t - x \sin(\omega t) \bigg) \dd t.
\end{align}
Using the relation $J_n (-x) = J_{-n}(x) = (-1)^n J_n(x)$, we get to $C_\text{WF}$ in Eq. \eqref{eq:C_WF}:
\begin{align}
	C_\text{WF} = - J_1 \left( \frac{a}{\omega} \right).
\end{align}
$C_\text{Syst.}$ which is given in Eq. \eqref{eq:C_Syst} becomes
\begin{align}
	C_\text{Syst.} &= \frac{1}{T} \int_0^T \cos(\omega t) \cos\bigg( \omega t + x \sin(\omega t) \bigg) \dd t \\
	&= \frac{1}{2T} \int_0^T \cos\bigg( 2 \omega t + x \sin(\omega t) \bigg) \dd t + \frac{1}{2T} \int_0^T \cos\bigg(x \sin(\omega t) \bigg) \dd t \\
	&= \frac{1}{2} \left[  J_0 \left( \frac{a}{\omega} \right) +  J_2 \left( \frac{a}{\omega} \right) \right],
\end{align}
which completes our derivation.

\section{Derivation of the Statistical Sensitivity} \label{app:Sensitivity}
A polarimeter measures the number of hit events recorded in the left, right, upper and lower sections, and the quantity of interest has either left-right asymmetry or upper-lower asymmetry, depending on the spin direction. For the vertical spin accumulation, we seek a signal from the left-right asymmetry, which is defined as
\begin{align}
	\epsilon(t) = \frac{L(t) - R(t)}{L(t) + R(t)} = P(t) A \theta(t),
\end{align}
where $L(t)$ and $R(t)$ are the counts in the left and right part of the polarimeter, respectively, $P(t) = P_0 e^{-t/\tau_p}$ is the spin polarization with a spin coherence time (SCT, $\tau_p$), $A$ is the analyzing power and $\theta(t) = \omega_d t$ is the vertical spin component. We want to determine the EDM signal $\omega_d$ with the smallest statistical uncertainty. The statistical uncertainty is determined by the $\chi^2$-minimizing fit normally to the regularly distributed statistics. But we study the case when the events are intentionally weighted, which can be beneficial in terms of the statistical uncertainty.

\begin{figure}[t]
	\centering
	\begin{subfigure}[t]{0.46\textwidth}
		\includegraphics[width=\textwidth]{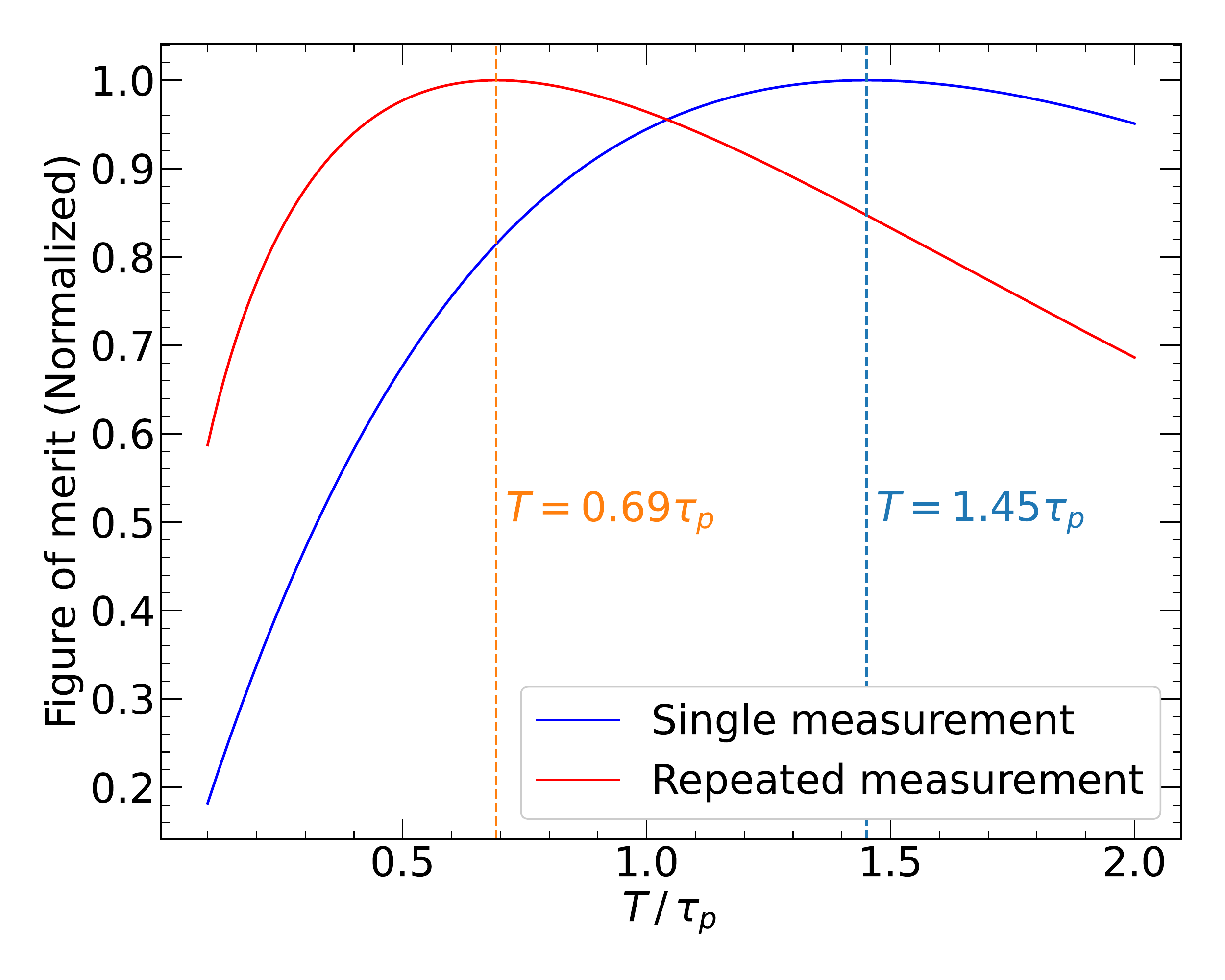}
		\caption{``linear fit'' method.}
		\label{fig:FOM_linearfit}
	\end{subfigure}
	~
	\begin{subfigure}[t]{0.46\textwidth}
		\includegraphics[width=\textwidth]{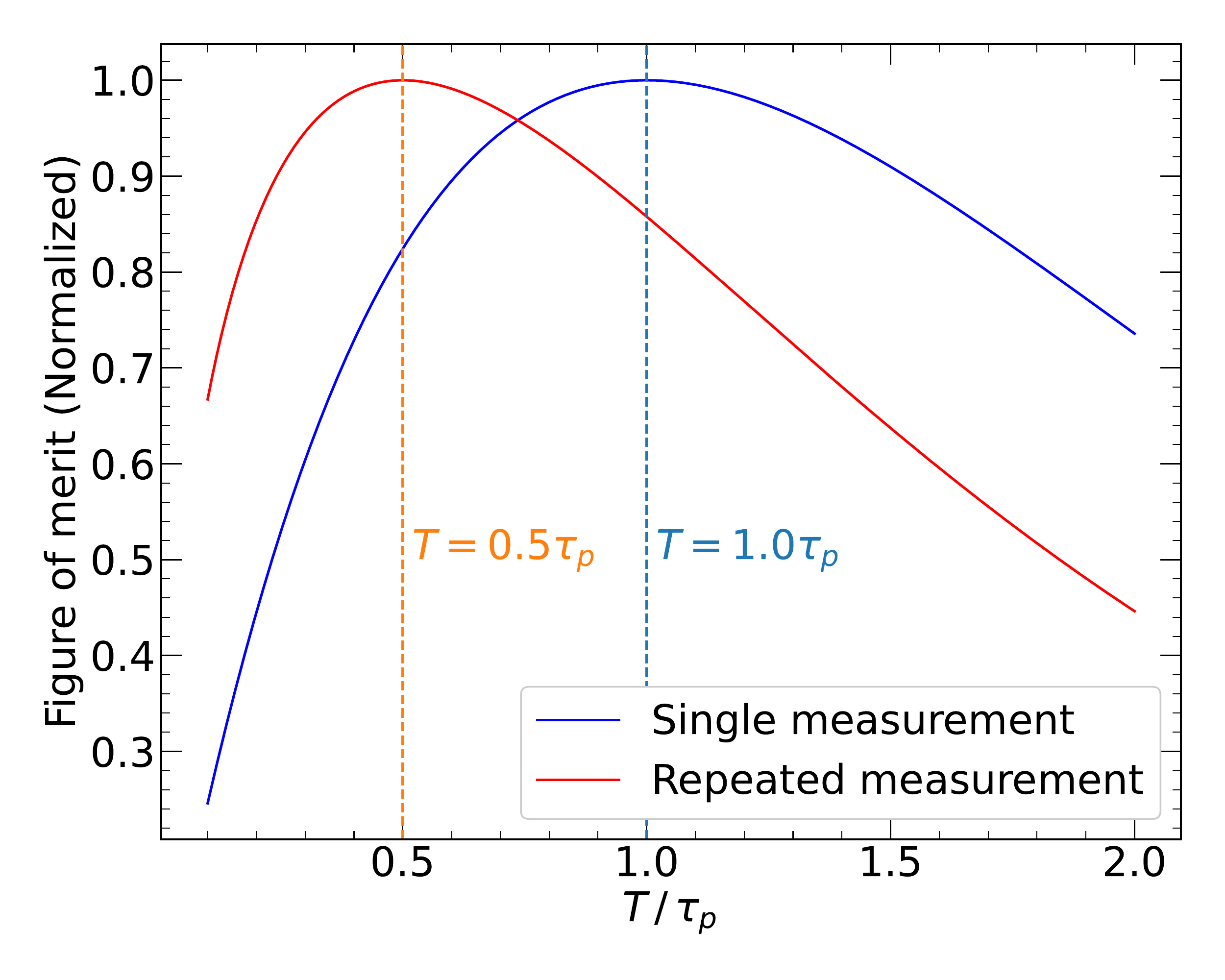}
		\caption{``take-all'' method.}
		\label{fig:FOM_takeall}
	\end{subfigure}
	\caption{Figure of merit for the statistical sensitivity, the inverse of the statistical uncertainty of $\omega_d$, as a function of the single measurement time $T$ divided by the spin coherence time $\tau_p$. Two methods are shown; ``linear fit'' determines the statistical uncertainty from $\chi^2$-minimization fit with a linear fit model, and ``take-all'' defines the slope by taking all data at a specific time $T$. The figure of merit is different for (blue) single measurement and (red) repeated measurement in a given experimental period.}
	\label{fig:}
\end{figure}

The $\chi^2$ for the model function $\alpha t$ fitting the polarimeter left-right asymmetry (just asymmetry hereafter) is given by\cite{EDMNote50}
\begin{align}
	\chi^2 = \sum_i \left( \frac{P_0 A \omega_d t_i e^{-t_i/\tau_p} - \alpha t_i}{\sigma_{\epsilon} (t_i)} \right)^2, \qquad \sigma_{\epsilon}^2 (t_i) = \frac{1 - \epsilon^2 (t_i)}{N(t_i)},
\end{align}
where $N(t_i) = L(t_i) + R(t_i)$. The corresponding uncertainty on the fit parameter $\alpha$ becomes
\begin{align} \label{eq:sigma_alpha_sqr}
	\sigma_\alpha^2 = \left( \frac{1}{2} \frac{\partial^2 \chi^2}{\partial \alpha^2} \right)^{-1} = \left( \sum_i \frac{t_i^2 N(t_i)}{1 - \epsilon(t_i)^2} \right)^{-1} \approx \left( \sum_i t_i^2 N(t_i) \right)^{-1},
\end{align}
where the last approximation holds because $\epsilon \ll 1 $. We have the total number of counts $N_\text{tot} = \sum_i N(t_i) \approx \int N(t)/b \; \dd t$ where $b$ is the bin width for each channel of the histogram. Let us denote $w(t)$ to be a measurement weighting function. If we take the polarimeter data stably and uniformly, then we have $w(t) = b/T$ where $T$ is the measurement time. From Eq. \eqref{eq:sigma_alpha_sqr}, one obtains
\begin{align}
	\sigma_\alpha^2 \approx \left( \int_0^T t^2 \frac{N_\text{tot}}{T} \right)^{-1}
	= \frac{3}{T^2 N_\text{tot}}.
\end{align}
To convert $\sigma_\alpha$ to $\sigma_{\omega_d}$, one needs to obtain the optimum fit parameter $\alpha$ by solving $\partial_\alpha \chi^2 = 0$, and the resulting $\alpha$ is
\begin{align}
	\alpha = 3 P_0 A \omega_d \frac{2\tau_p^3 - \tau_p \left[ (T + \tau_p)^2 + \tau_p^2 \right] e^{-T/\tau_p}}{T^3}.
\end{align}
Hence, it follows that
\begin{align} \label{eq:sigma_omega_d}
	\sigma_{\omega_d} = \frac{1}{P_0 A \sqrt{3 N_\text{tot}}} \; \frac{T^2}{2\tau_p^3 - \tau_p \left[ (T + \tau_p)^2 + \tau_p^2 \right] e^{-T/\tau_p}}.
\end{align}
Numerically, the optimum measurement time $T$ that minimizes $\sigma_{\omega_d}$ is $T \approx 1.45 \tau_p$, but it is only 5\% different when $T = \tau_p$ and obviously it is not so economic to have a 44\% longer measurement to decrease the statistical uncertainty by 5\%. In general, it is reasonable to minimize the statistical uncertainty of a \textit{repeated measurement} rather than a single measurement. Let $T_\text{exp}$ be the total experimental period, then we convert $N_\text{tot}$ into
\begin{align}
	N_\text{tot} \rightarrow N_\text{cyc} \frac{T_\text{exp}}{T},
\end{align}
where $N_\text{cyc}$ is the number of stored particles for a single measurement (one cycle). Then the optimum $T$ which minimizes $\sigma_{\omega_d}$ becomes $T \approx 0.69 \tau_p$. Defining a figure of merit as the inverse of $\sigma_{\omega_d}$, its dependence on the single measurement time $T$ is given in Fig. \ref{fig:FOM_linearfit}.
Substituting $T \approx 0.69 \tau_p$ into Eq. \eqref{eq:sigma_omega_d}, we obtain
\begin{align} \label{eq:sigma_omega_d_linearfit_repeated}
	\sigma_{\omega_d} \approx \frac{3.47}{P_0 A \sqrt{N_\text{cyc} T_\text{exp} \tau_p}}.
\end{align}

This statistical uncertainty can be reduced by up to nearly 30\% by weighting the later statistics more than the earlier ones. As an extreme limit, one can imagine taking all particles at a given time $T$, and finding the slope from $\epsilon(T)/T$ instead of fitting:
\begin{align}
	\frac{\epsilon(T)}{T} = P_0 A \omega_d e^{-T/\tau_p}.
\end{align}
Then $\sigma_{\omega_d}$ is now given by $\sigma_\epsilon$,
\begin{align} \label{eq:sigma_omega_d_take_all}
	\sigma_{\omega_d} = \frac{\sigma_\epsilon}{P_0 A T e^{-T/\tau_p}}
	\approx \frac{1}{P_0 A T e^{-T/\tau_p} \sqrt{N_\text{tot}}}.
\end{align}
This is minimized when $T = \tau_p$. Then considering the repeated measurement in the total experimental tim $T_\text{exp}$, Eq. \eqref{eq:sigma_omega_d_take_all} becomes
\begin{align}
	\sigma_{\omega_d} = \frac{1}{P_0 A e^{-T/\tau_p} \sqrt{N_\text{cyc} T_\text{exp} T}},
\end{align}
which now is minimized when $T = \tau_p/2$, yielding
\begin{align} \label{eq:sigma_omega_d_takeall_repeated}
	\sigma_{\omega_d} \approx \frac{2.33}{P_0 A \sqrt{N_\text{cyc} T_\text{exp} \tau_p}}.
\end{align}
The figure of merit for this take-all method is shown in Fig. \ref{fig:FOM_takeall}. Comparing Eqs. \eqref{eq:sigma_omega_d_linearfit_repeated} and \eqref{eq:sigma_omega_d_takeall_repeated}, the sensitivity of the take-all method is smaller than that of the linear fit method by a factor of $3.47 / 2.33 \approx 1.5$. Practically, the intensity of the stored beam is limited by intra-beam-scattering, which depends on the intensity of the stored beam. Therefore, a large intensity beam can be injected and used (extracted on the polarimeter target) very quickly to determine the vertical spin component of the beam at early times with high accuracy. The rest of the beam can be taken out at the end of the storage time to again determine with high accuracy. That would be, statistically, the most sensitive method for determining the vertical spin precession rate. However, for high sensitivity, a small part of the beam would also need to be used to make sure the longitudinal spin component is kept along with the momentum.

\begin{table}[t]
	\centering
	\caption{The statistical uncertainty of the vertical spin precession rate ($\sigma_{\omega_d}$) which is minimized by the optimum single measurement time $T$ for different configurations. The linear fit with no weight is a method to take data uniformly and continuously during storage, while the take-all method measures all particles at a specific time, which is an extreme case of giving weights for the measurement.}
	\label{tab:sigma_omega_d}
	\begin{tabular}{lcc}
		\hline\hline
		Method & Single measurement & Repeated measurement for $T_\text{exp}$ \\
		\hline \\[-2ex]
		Linear fit (no weight) & $\sigma_{\omega_d} (T = 1.45 \tau_p) = \dfrac{3.40}{P_0 A \tau_p \sqrt{\kappa N_\text{tot}}}$ & $\sigma_{\omega_d} (T = 0.69 \tau_p) = \dfrac{3.47}{P_0 A \sqrt{\kappa N_\text{cyc} T_\text{exp} \tau_p}}$ \\[4ex]
		Take-all (100\% at latest time) & $\sigma_{\omega_d} (T = \tau_p) \phantom{1.00} = \dfrac{2.72}{P_0 A \tau_p \sqrt{\kappa N_\text{tot}}}$ & $\sigma_{\omega_d} (T = 0.5 \tau_p) \phantom{0} = \dfrac{2.33}{P_0 A \sqrt{\kappa N_\text{tot} T_\text{exp} \tau_p}}$ \\[3ex]
		\hline\hline
	\end{tabular}
\end{table}

The optimum sensitivities of the EDM signal, $\sigma_{\omega_d}$, for different configurations and methods are summarized in Table \ref{tab:sigma_omega_d}, with the optimum single measurement time. In this paper, we use the best sensitivity obtained by the take-all method with repeated measurement shown in Eq. \eqref{eq:sigma_omega_d_takeall_repeated} to estimate the statistical sensitivity in Sec. \ref{sec:sensitivity}. To convert $\sigma_{\omega_d}$ to the sensitivity of the EDM directly, we recall the relation between the vertical spin slope and the EDM, as in the last column of Table \ref{tab:various_methods}:
\begin{align}
	\omega_d = \frac{d_\text{AC}}{2s\hbar} E^* C_\text{WF}.
\end{align}
Accordingly, we arrive to the final sensitivity expression for the AC EDM ($\sigma_d$):
\begin{align}
	\sigma_d = \frac{4.67 s \hbar}{P_0 A E^* C_\text{WF} \sqrt{\kappa N_\text{cyc} T_\text{exp} \tau_p}},
\end{align}
where $\kappa$ is the detector efficiency so that the number of polarimeter events is proportional to $\kappa$.

\twocolumngrid
\bibliography{AxionEDM_RFWF}

\end{document}